\documentclass[conference]{IEEEtran}
\IEEEoverridecommandlockouts
\usepackage{cite}
\usepackage{amsmath,amssymb,amsfonts}
\usepackage{algorithmic}
\usepackage{graphicx}
\usepackage{textcomp}
\usepackage{xcolor}
\usepackage{url}
\usepackage[colorlinks=true,linkcolor=black,citecolor=black,urlcolor=blue]{hyperref}
\usepackage{booktabs}
\usepackage{array}
\usepackage{multirow}

\usepackage{cuted} 
\usepackage{capt-of}

\def\BibTeX{{\rm B\kern-.05em{\sc i\kern-.025em b}\kern-.08em
    T\kern-.1667em\lower.7ex\hbox{E}\kern-.125emX}}
\makeatletter
\renewcommand\footnoterule{%
  \kern -3pt
  \hrule width \columnwidth
  \kern 2.6pt
}
\makeatother

\begin{document}

\title{Auto-Slides: An Interactive Multi-Agent System for Creating and Customizing Research Presentations}

\author{
Yuheng Yang$^{1}$,
Wenjia Jiang$^{1}$,
Yang Wang$^{1}$,
Yi Song$^{2}$,
Yiwei Wang$^{3}$,
Chi Zhang$^{1,*}$\thanks{*Corresponding author.}\\
$^{1}$AGI Lab, Westlake University \quad
$^{2}$Teeni AI \quad
$^{3}$University of California at Merced\\
yangyuheng@westlake.edu.cn, chizhang@westlake.edu.cn\\
Project Page: \href{https://auto-slides.github.io/}{https://auto-slides.github.io/}
}

\maketitle

\begin{strip}
    \vspace{-1.85cm}
    \centering
    
    \includegraphics[width=0.97\textwidth]{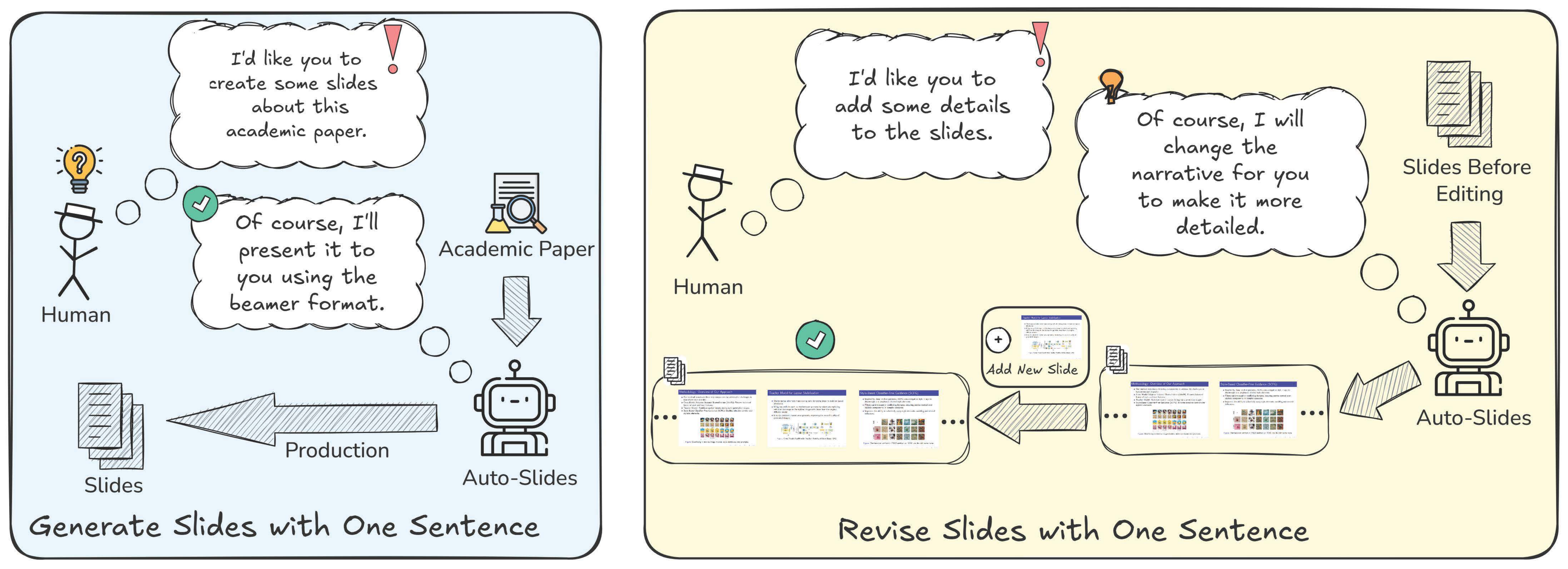}
    \captionof{figure}{\textbf{\textsc{Auto-Slides}' capabilities.} \emph{Left:} Users can generate complete academic presentation slides from an academic paper. \emph{Right:} Users can iteratively revise the generated slides by providing high-level natural language instructions, enabling efficient and precise slide editing.}
    \label{fig:teaser}
    \vspace{-0.45cm}
\end{strip}

\begin{abstract}

The rapid progress of large language models (LLMs) has opened new opportunities for education. While learners can interact with academic papers through LLM-powered dialogue, limitations still exist: the lack of structured organization and the heavy reliance on text can impede systematic understanding and engagement with complex concepts. To address these challenges, we propose \textsc{Auto-Slides}, an LLM-driven system that converts research papers into pedagogically structured, multimodal slides (\emph{e.g.}, diagrams and tables). Drawing on cognitive science, it creates a presentation-oriented narrative and allows iterative refinement via an interactive editor to better match learners' knowledge level and goals. \textsc{Auto-Slides} further incorporates verification and knowledge retrieval mechanisms to ensure accuracy and contextual completeness. Through extensive user studies, \textsc{Auto-Slides} demonstrates strong learner acceptance, improved structural support for understanding, and expert-validated gains in narrative quality compared with conventional LLM-based reading. Our contributions lie in designing a multi-agent framework for transforming academic papers into pedagogically optimized slides and introducing interactive customization for personalized learning.

\end{abstract}

\begin{IEEEkeywords}
Computing education, Human-computer interaction, Information visualization
\end{IEEEkeywords}



\section{Introduction}

In recent years, the rapid advancement of large language models (LLMs) has significantly transformed the landscape of natural language processing and artificial intelligence. Models such as GPT\cite{brown2020language}, PaLM\cite{chowdhery2022palm}, and LLaMA\cite{touvron2023llama} have demonstrated remarkable capabilities in understanding, generating, and reasoning with human language, enabling a wide range of applications in information retrieval, automated writing, programming assistance, and beyond. With their growing accessibility through public APIs and interactive interfaces, LLMs are increasingly integrated into both professional and educational contexts, reshaping how individuals acquire, process, and apply information.

Alongside these technological developments, the integration of LLMs into education and learning has emerged as a notable trend. Unlike traditional learning methods, where students passively consume materials such as textbooks or academic papers, LLM-based systems allow learners to interact directly with content in a dynamic and dialogic manner. For example, instead of merely reading an academic paper, a learner can upload the document to an LLM-powered system and engage in iterative questioning and clarification. Through this interactive process, learners are able to query definitions, verify interpretations, and explore related concepts in real time. This paradigm shift holds the potential to make complex academic knowledge more approachable, supporting personalized exploration, fostering deeper comprehension, and reducing barriers to engaging with highly specialized content.

However, directly learning from LLM-based interactions still presents notable limitations. First, the knowledge acquired through dialogue often lacks structure. Academic papers, for example, are carefully organized into sections that convey background, motivation, methodology, experiments, and conclusions in a coherent framework. When learners interact with LLMs through isolated questions and answers, they primarily focus on specific details while losing sight of the overall structure of the work. This piecemeal approach makes it difficult to form a systematic understanding of the material. Second, the output modality of current LLM-based learning is predominantly textual. While text-based explanations are useful, they fall short in domains where comprehension is enhanced by multimodal representations, such as diagrams, tables, and visual summaries. The absence of such elements can hinder learners' ability to grasp complex ideas and relationships.
These limitations highlight the need for systems that can leverage LLMs to present knowledge from academic materials in a structured, learner-friendly, and multimodal format. In particular, we focus on learning from research papers, where both systematic organization and visual representation play critical roles. To address this need, we propose \textsc{Auto-Slides}, an LLM-driven system that automatically transforms research papers into visually rich, well-structured, and pedagogically optimized slide decks, thereby supporting more effective and engaging learning experiences overall.

Transforming a research paper into effective presentation slides poses several challenges. First, there is a fundamental gap between the narrative style of academic writing and the pedagogical logic required for teaching and learning. Research papers are typically written to document novelty and rigor, often assuming that readers already possess relevant background knowledge. This structure, while suitable for scholarly communication, can be difficult for learners to follow and may obscure key ideas. Educational psychology provides evidence that careful content reorganization can significantly enhance learning effectiveness. For example, Cognitive Load Theory\cite{sweller1988cognitive} suggests that instructional materials should be arranged to reduce extraneous cognitive load, enabling learners to focus mental resources on essential understanding. Mayer's Multimedia Learning Theory\cite{mayer2009multimedia} further emphasizes that presenting information in a coherent sequence, and combining verbal and visual forms, facilitates comprehension and knowledge retention. Drawing on these principles, we design a content reorganization agent that converts the logical flow of a paper into a presentation-oriented narrative. Specifically, we adopt a Problem–Method–Result–Conclusion (PMRC) framework to restructure a paper's background, motivation, methodology, and findings into a format that is easier to follow, thereby helping learners comprehend and internalize the material.

Second, effective learning requires adaptability to diverse learner backgrounds and goals. A static set of slides may not be sufficient, as learners differ in their prior knowledge, interests, and the level of detail they require. To overcome this limitation, we introduce an interactive editor agent that supports iterative refinement of the automatically generated slides. Through natural language dialogue, users can request modifications—for example, adding additional slides on specific concepts or simplifying technical details. This interactive capability enables the system to dynamically tailor slide decks to individual learners, ensuring that the generated materials are not only accurate and coherent but also aligned with the user's unique learning needs.

Finally, beyond the content reorganization and interactive editing capabilities introduced above, \textsc{Auto-Slides} incorporates two additional mechanisms to further enhance the reliability and educational value of the generated slides. First, a Verification–Adjustment Loop compares the slide content against the source paper to identify and repair omissions or inaccuracies, ensuring complete and faithful coverage of key contributions. Second, an external knowledge retrieval module augments the Editor Agent, enabling it to enrich slides with background information from cited or related works when users request clarification of unfamiliar concepts. Guided by overarching principles from educational psychology and cognitive science, these complementary components strengthen \textsc{Auto-Slides}'s ability to deliver accurate, comprehensive, and contextually enriched learning materials.

We conduct multiple user studies to evaluate \textsc{Auto-Slides} from different perspectives, including the overall quality of the generated slides, the usefulness of the interactive customization function, and the comparative advantages of using \textsc{Auto-Slides} versus directly reading papers with LLM assistance. The results consistently show that \textsc{Auto-Slides} provides stronger structural support for understanding complex academic content while improving learner experience and expert-rated narrative quality.
In summary, this paper makes the following key contributions: 1) We design a novel system that automatically transforms academic papers into structured, visually enriched, and pedagogically informed slide decks, drawing on principles from educational psychology and cognitive science. 2) We introduce an editor agent that supports dynamic and dialogic refinement of slides, allowing learners to personalize content according to their knowledge background, learning goals, and preferences.
\section{Method}
\label{sec:method}

\begin{figure*}[t]
    \centering
    \includegraphics[width=1\linewidth]{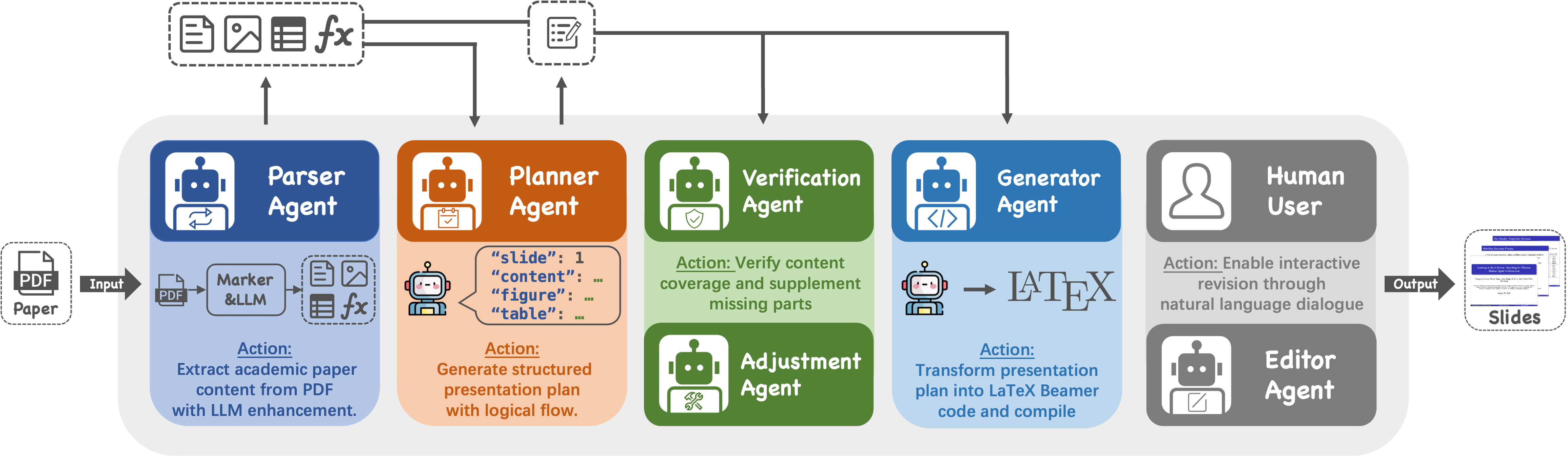}
    \caption{\textbf{Overview of \textsc{Auto-Slides}.} The multi-agent pipeline transforms papers into slides via three stages: (1) \textbf{Content Understanding}, where Parser and Planner Agents design the slide structure in JSON; (2) \textbf{Quality Assurance}, utilizing Verification and Adjustment Agents to ensure content fidelity; and (3) \textbf{Generation \& Interaction}, where Generator and Editor Agents produce LaTeX slides and facilitate human-in-the-loop revisions.}
    \label{fig:method}
\end{figure*}

As illustrated in Fig.~\ref{fig:method}, \textsc{Auto-Slides} orchestrates a multi-agent pipeline comprising three key modules: the \textbf{Parser} and \textbf{Planner Agents} for content understanding and structural design; the \textbf{Verification–Adjustment Agents} for ensuring factual accuracy and completeness; and the \textbf{Generator} and \textbf{Editor Agents}, which produce multi-modal slides and enable human-in-the-loop refinement.

\subsection{Content Understanding and Structuring}
This initial phase addresses the dual challenges of extracting multi-modal information from structurally complex PDFs and re-conceptualizing the narrative for presentation.

\paragraph{High-Fidelity Multi-modal Parsing.}
To address the complexity of academic PDFs, our Parser Agent employs a two-stage strategy. First, it utilizes the Marker\cite{marker} model to convert the document into a high-fidelity structured Markdown format. Second, to prevent context truncation and structural errors in downstream processing, an LLM explicitly extracts table and equation blocks from this Markdown, preserving their original syntax in a structured JSON file (see Appendix). This allows subsequent agents to retrieve complex multi-modal data accurately without scanning the full text.

\paragraph{Cognitive-Driven Presentation Planning.}
The Planner Agent transforms extracted content from reader-centric IMRaD\cite{day2012how} to presenter-centric Problem--Motivation--Results--Conclusion (PMRC) narrative. Guided by Cognitive Load Theory\cite{sweller1988cognitive,sweller2011cognitive} and Multimedia Learning Principles\cite{mayer2009multimedia}, the agent reduces extraneous load by sequencing concepts incrementally and optimizing spatial contiguity between text and visuals. The final output is a comprehensive JSON plan containing the sequenced narrative, bullet points, and visual aids.

\subsection{Quality Assurance and Content Adjustment}
Although LLM-based planning effectively structures slide content, LLMs still have inherent limitations such as context length constraints, where earlier information may be omitted, and hallucination tendencies, where generated content can deviate from source facts. In automated slide generation, these issues may omit important findings or introduce factual inaccuracies in the generated slides.

To address this, we introduce a two-stage quality assurance step involving a Verification Agent followed by an Adjustment Agent when needed. The Verification Agent calls an LLM to compare the generated slide plan against the structured content extracted from the original paper, focusing on methodology, results, and conclusions. Instead of requiring exact sentence-level matches, it uses a loose semantic criterion in which coverage is counted when the essential concept or finding is explicitly conveyed. It then assesses each content area for coverage sufficiency; only when overall coverage is clearly insufficient or high-importance items are missing does it flag the plan for revision. In that case, the Adjustment Agent is triggered to repair only the high-importance omissions. This involves locating the relevant material in the original paper, integrating it into existing slides where possible, or creating a new supplemental slide if necessary. The LLM reformulates the retrieved content into 2--4 concise bullet points per slide, written in professional academic presentation style to ensure clarity and consistency.

By inserting this two-stage QA step between planning and slide generation, we prevent the propagation of factual errors, ensure that all high-value content from the source paper is represented, and maintain both the completeness and reliability of the automated presentation.

\subsection{Multi-modal Generation}
The preceding steps produce a structured slide plan in JSON format—either directly from the Planner Agent or, if necessary, after refinement through the Verification–Adjustment process. To transform this abstract plan into an actual multi-modal slide deck containing text, figures, tables, and equations, we generate \LaTeX{} source code and compile it into PDF slides. We adopt \texttt{Beamer}\cite{beamer} as the presentation framework due to its precise typography, native mathematical notation support, and flexible layout control, which make it particularly suitable for academic contexts. The \textbf{Generator Agent} parses the JSON blueprint and, using a user-supplied \texttt{Beamer}\cite{beamer} theme (where the theme specifies the color scheme and all other style parameters follow its defaults), calls an LLM to produce \LaTeX{} code for each slide: converting textual content into bullet points or paragraph blocks, embedding and positioning figures according to the theme's layout rules, reconstructing tables—often provided as Markdown—into \texttt{tabular} environments, and preserving any mathematical expressions in their original form. The generated \LaTeX{} source is compiled to PDF, and if compilation fails, an automatic correction routine captures the \texttt{.log} file and the faulty source, passes them to the LLM for targeted fixes, and re-compiles until successful or a preset attempt limit is reached. The final output is a complete, visually consistent academic presentation faithfully implementing the verified slide plan.
\begin{figure}[t]
    \centering
    \includegraphics[width=1\linewidth]{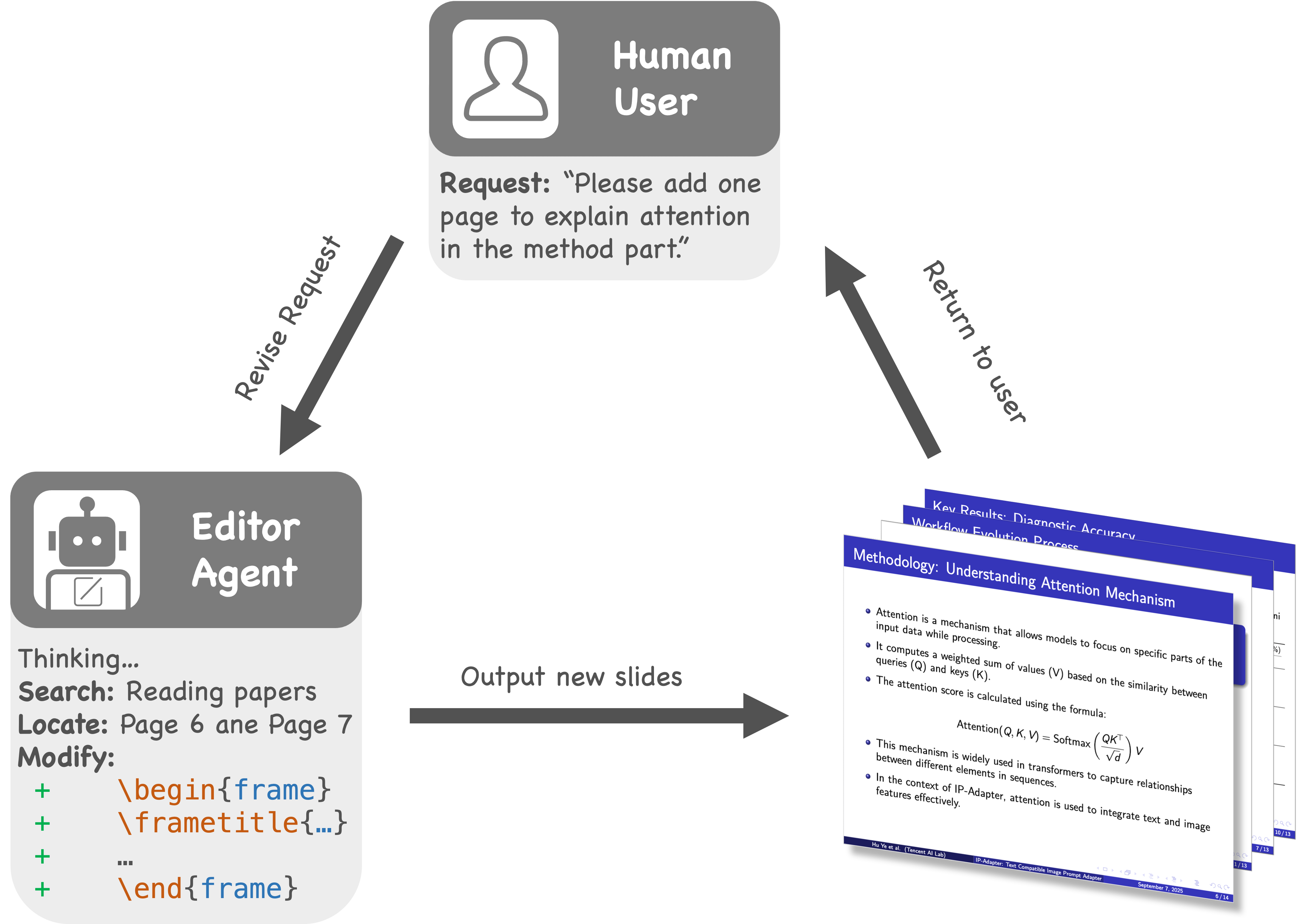}
    \caption{\textbf{Interactive optimization workflow.} The \textbf{Editor Agent} processes natural language commands via a ReAct\cite{yao2023react} pipeline, orchestrating content retrieval and LaTeX code modifications to compile updated presentations.}
    \label{fig:interactive}
\end{figure}

\subsection{Interactive Optimization}
To adapt content to diverse learner backgrounds, we introduce the \textbf{Editor Agent}, which enables on-demand refinement via natural language commands. Operating within a ReAct\cite{yao2023react} framework, the agent decomposes user requests into a sequence of atomic actions, specifically \texttt{locate}, \texttt{search}, \texttt{modify}, \texttt{insert}, and \texttt{delete}, to dynamically edit the underlying \LaTeX{} source code. This allows the system to not only adjust existing text but also retrieve external knowledge from cited papers to explain unfamiliar concepts. Figure~\ref{fig:interactive} illustrates this workflow, where the agent orchestrates these tools to iteratively compile updated slide decks. Detailed definitions of these operation primitives and the retrieval mechanism are provided in the Supplementary Material.

\section{Experiments}
\label{sec:experiments}
We evaluate \textsc{Auto-Slides} through three user studies and one automated ablation, assessing: (i) the impact of interactive editing on learner engagement (Section~\ref{sec:learner_study}); (ii) slide-based versus chat-based learning (Section~\ref{sec:comparative_study}); (iii) the benefits of narrative optimization (Section~\ref{sec:expert_study}); and (iv) component-level contributions to content fidelity (Section~\ref{sec:llm_as_judge}). Detailed implementation settings and statistical methods are provided in the Supplementary Material.
\subsection{Evaluation Metrics}
We employed a multi-faceted evaluation protocol involving learners, domain experts, and LLM judges.

\textbf{Human Evaluation (Learners \& Experts).} 
All human participants rated items on 7-point Likert scales (1=Strongly Disagree, 7=Strongly Agree), which were averaged into composite scores. Full instruments are detailed in Appendices.
\begin{itemize}
    \item \emph{Learner Interactive Perception:} Assessed \emph{Learning Enhancement} (concept support) and \emph{Control \& Agency} (autonomy).
    \item \emph{Learner Comparative Experience:} Contrasted Slide \emph{vs.} Chat formats across five dimensions: \emph{Visual Clarity}, \emph{Structural Organization}, \emph{Learning Flow}, \emph{Understanding Support}, and \emph{Overall Satisfaction}.
    \item \emph{Expert Assessment:} Evaluated narrative quality on \emph{Content Accuracy} (factual correctness), \emph{Information Density} (content balance), and \emph{Narrative Structure \& Flow} (logical progression).
\end{itemize}

\textbf{LLM Comparative Evaluation.} 
Judge models performed blind pairwise comparisons on anonymized slide pairs using a five-point scale (subsequently collapsed to Win/Tie/Loss). To mitigate position bias, we adopted a bidirectional protocol (evaluating both \textbf{forward} and \textbf{reverse} orders), recording the \emph{Forward--Reverse Agreement} (FRA) to quantify reliability and ensure consistent aggregation.

\subsection{User Study 1: Interactive Slide Functions from Learners' Perspective}
\label{sec:learner_study}


\textbf{Goal and Participants.} To evaluate how interactive customization affects learners' comprehension, engagement, and sense of control when studying unfamiliar research topics, we recruited 30 undergraduate students ($N=30$) from diverse academic backgrounds. Participants acted as typical audience members rather than content creators. Detailed demographics are provided in the Supplementary Material.

\textbf{Materials.}
The learning material was drawn from a recently published, award‑winning HCI research paper\cite{cucumak2025designing} on urban community–animal interactions, chosen for its strong narrative structure and minimal reliance on domain‑specific technical knowledge, in order to reduce comprehension barriers across different academic backgrounds.
An initial slide deck was automatically generated from the paper using the \textsc{Auto-Slides} workflow described in Section~\ref{sec:method}, providing concise textual summaries alongside key figures and diagrams.
We also enabled the interactive editing and augmentation functions in the \textsc{Auto-Slides} interface, allowing participants to add, remove, or elaborate on slide content according to their learning needs throughout each study session.

\textbf{Procedure.}
Participants first read through the automatically generated initial slide deck at their own pace. 
They then used the interactive slide editing functions to modify and expand the deck until they felt they had sufficiently understood the paper's content. 
No strict time limit was imposed, but most sessions lasted between 15 and 25 minutes in total.  
Interaction logs (\emph{e.g.}, number of edits, added elements) were recorded for exploratory analysis, though the primary focus of this study was participants' subjective perceptions.
Immediately after completing the interactive learning session, participants filled out a 7-point Likert-scale questionnaire (1 = Strongly Disagree, 7 = Strongly Agree) assessing two key dimensions: \textit{Learning Enhancement} (\emph{e.g.}, \emph{``The interactive functions made it easier for me to understand the paper''}) and \textit{Control and Agency} (\emph{e.g.}, \emph{``I felt in control of how I viewed and organized the paper's content''}). 
Each dimension was measured by multiple items (full wording in Appendix); internal consistency was acceptable for both scales ($\alpha > .80$).
We conceptualized this study as an exploratory single-condition evaluation aimed at gauging subjective acceptance and perceived utility, without a non-interactive control condition.


\begin{table}[t]
\centering
\caption{Results of \textbf{User Study~1} (Interactive Functions). Mean ratings on a 7-point scale compared against the neutral midpoint of 4 using one-sample $t$-tests.}
\begin{tabular}{lccccc}
\toprule
\textbf{Dimension} & \textbf{M} & \textbf{SD} & \textbf{$t$(df)} & \textbf{$p$} & \textbf{$d$} \\
\midrule
Learning Enhancement & 5.46 & 1.12 & 7.13 (29) & $<.001$ & 1.30 \\
Control and Agency   & 5.49 & 1.16 & 7.02 (29) & $<.001$ & 1.28 \\
\bottomrule
\end{tabular}
\label{tab:learner_summary}
\vspace{-0.6cm}
\end{table}
\textbf{Results.} 
Participants rated interactive functions significantly above the neutral midpoint in both \emph{Learning Enhancement} and \emph{Control and Agency} (Table~\ref{tab:learner_summary}). Qualitative feedback reinforced this, with users noting that the tools helped them \emph{``focus on key points''} and \emph{``organize information,''} though a minority cautioned against potential distraction. Ultimately, the capacity to actively adjust content appears to foster a sense of ownership and agency, reinforcing engagement and facilitating deeper cognitive processing of the material.

\subsection{User Study 2: Comparing Slides and LLM Chat-based Learning from Learners' Perspective}
\label{sec:comparative_study}


\textbf{Goal and Participants.} To benchmark \textsc{Auto-Slides} against standard LLM chat interfaces, we recruited 24 researchers ($N=24$) with backgrounds in HCI and Computer Science, balanced by prior LLM proficiency to evaluate relative strengths across modalities.

\textbf{Materials.} 
We prepared three peer-reviewed HCI papers spanning diverse topics and structures (\emph{e.g.}, empirical study, system design, theoretical framing).  
All papers were unfamiliar to participants, as confirmed during screening.  
Using \textsc{Auto-Slides}, we automatically generated slide decks for each paper, summarizing motivation, methods, results, and conclusions with bullet points, key figures, and diagrams. 
For the LLM chat condition, we used GPT-4\cite{openai2023gpt4} via ChatGPT with a standardized prompt:  
\textit{``Summarize the attached paper for efficient learning including key points, main findings, and relevant figures. Answer follow-up questions from the user.''}  
The model received the full paper in PDF form, and the conversation history contained no extraneous context.  
No manual post-editing was performed in either condition. 

\textbf{Procedure.} 
We employed a counterbalanced within-subject design where participants engaged with both the \textit{Slide} (free navigation) and \textit{Chat} (LLM interaction with open follow-up) conditions for up to 15 minutes each. Evaluation followed a two-stage protocol: 
(1) \textit{Absolute Ratings}, a 7-point Likert survey on five dimensions (visual/structural clarity, understanding, engagement, satisfaction) administered immediately post-condition; and 
(2) \textit{Comparative Ratings} (see Supplementary Material), which directly contrasted the two formats across the same dimensions upon study completion. 
Finally, participants indicated their overall preference via a forced-choice question and provided open-ended feedback.

\begin{table}[t]
\centering
\footnotesize 
\renewcommand{\arraystretch}{1.25} 

\caption{Results of \textbf{User Study~2} (Slides \emph{vs.}\ LLM Chat). Values are Mean (SD) on a 7-point Likert scale. Significance tested via paired $t$-tests with Holm–Bonferroni correction. \textbf{Abbreviations:} VCI: Visual clarity \& intuitiveness; SCO: Structural clarity \& organization; SUM: Support for understanding \& memory; LFE: Learning flow \& engagement; OS: Overall satisfaction.}

\begin{tabular*}{\columnwidth}{@{\extracolsep{\fill}} l ccccc @{}}
\toprule
\textbf{Dim.} & 
\textbf{\shortstack{Slides\\M (SD)}} & 
\textbf{\shortstack{Chat\\M (SD)}} & 
\textbf{$t(23)$} & 
\textbf{$p_{\mathrm{adj}}$} & 
\textbf{$d_z$} \\
\midrule
VCI & 6.10 (0.56) & 5.05 (0.64) & 6.21 & $<.001^{***}$ & 1.27 \\
SCO & 5.90 (0.60) & 5.00 (0.65) & 5.38 & $<.001^{***}$ & 1.10 \\
SUM & 5.50 (0.62) & 5.10 (0.68) & 2.92 & $<.05^{*}$    & 0.60 \\
LFE & 5.15 (0.70) & 5.05 (0.75) & 0.51 & .616          & 0.10 \\
OS  & 5.25 (0.68) & 5.10 (0.73) & 1.02 & .318          & 0.21 \\
\bottomrule
\end{tabular*}
\label{tab:comparative_absolute_summary}
\end{table}

\textbf{Results.} 
Quantitative metrics (Table~\ref{tab:comparative_absolute_summary}) and post-study ratings consistently show that the slide-based format outperformed standard LLM chat in visual clarity, structural organization, and support for understanding. Consequently, most participants expressed an overall preference for the slide format. Qualitative feedback elucidated a complementary relationship: participants valued slides for their \textit{visual and structural grounding}, which provides a necessary scaffold often missing in open-ended chat, while appreciating the \textit{interactive exploration} offered by LLMs. These findings support a \textit{hybrid workflow} where slides ensure essential content coverage, laying a structured foundation for subsequent personalized inquiry via LLM chat.

\subsection{User Study 3: Evaluating Narrative Structure Optimization from Experts' Perspective}
\label{sec:expert_study}
\textbf{Goal and Participants.} To investigate whether narrative optimization (restructuring IMRaD\cite{day2012how} content into a PMRC flow) improves slide quality, we recruited 8 experienced researchers ($N=8$). Participants evaluated slides generated from their own papers to leverage deep subject familiarity, comparing the full system against an ablated baseline on dimensions of accuracy, information density, and narrative flow in presentation.


\textbf{Materials.} 
We generated fully automated slide decks for the expert papers using two system configurations: 
(1) \emph{With Narrative Optimization} (Full Pipeline), where the \texttt{Planner Agent} restructures the original IMRaD\cite{day2012how} content into a pedagogical Problem--Motivation--Results--Conclusion (PMRC) narrative guided by cognitive principles\cite{sweller1988cognitive,mayer2009multimedia,paivio1986mental}; and 
(2) \emph{Without Narrative Optimization} (Ablation), which preserves the paper's original linear sequence. 
All other pipeline stages (parsing, verification) remained identical, and no human post-editing was performed.

\textbf{Procedure.} 
For each participant, one of their papers was randomly assigned to the ``with optimization'' configuration and the other to the ``without optimization'' configuration, with this assignment counterbalanced across participants to control for paper‑specific effects. 
The order in which the two slide decks were presented was randomized to mitigate order effects. 
Participants reviewed each deck in turn and then completed the 9‑item rubric for that deck. Ratings were entered directly into a digital form.

\begin{table}[t]
\centering
\small 
\renewcommand{\arraystretch}{1.25} 


\caption{Results of \textbf{User Study~3} ($n=8$). Expert evaluation of slides with \emph{vs.} without narrative structure optimization. Values are mean ratings (7-point Likert scale) compared via paired $t$-tests.}

\begin{tabular*}{\columnwidth}{@{\extracolsep{\fill}} l ccccc @{}}
\toprule
\textbf{Dimension} & 
\textbf{\shortstack{w/ Opt.\\M (SD)}} & 
\textbf{\shortstack{w/o Opt.\\M (SD)}} & 
\textbf{$t(7)$} & 
\textbf{$p$} & 
\textbf{$d_z$} \\
\midrule
Content accuracy & 5.59 (1.05) & 4.96 (0.92) & 2.80 & .026 & 0.99 \\
Info. density    & 4.38 (1.12) & 4.50 (1.25) & 0.40 & .704 & 0.14 \\
Narrative flow   & 4.96 (1.08) & 4.30 (1.16) & 3.03 & .019 & 1.07 \\
\bottomrule
\end{tabular*}
\label{tab:expert_summary}
\vspace{-0.6cm}
\end{table}
\textbf{Results.} 
Slides incorporating narrative structure optimization outperformed ablation baselines in both \emph{Content Accuracy} and \emph{Narrative Structure \& Flow}, while \emph{Information Density} remained consistent (Table~\ref{tab:expert_summary}). Experts attributed these gains to smoother conceptual navigation and higher fidelity in presenting findings, indicating that the method enhances understanding by reorganizing content rather than altering its volume. Ultimately, these narrative improvements reduce cognitive load, making it easier for audiences to follow complex arguments and verify evidence in dense presentations.

\subsection{Automatic Evaluation: LLM-as-Judge Component Ablation}
\label{sec:llm_as_judge}


\textbf{Goal.} To evaluate the contribution of the Parser Agent to complex element fidelity and the Verification--Adjustment Agents to factual accuracy, we conducted ablation studies on 14 diverse academic papers. We compared the \textbf{Full} \textsc{Auto‑Slides} pipeline against two variants: \textbf{w/o InfoExt} (omitting second-stage parsing) to assess table fidelity, and \textbf{w/o VerifLoop} (omitting verification agents) to isolate content accuracy improvements. For consistency, generated \LaTeX{} slides were normalized into a layout-agnostic \texttt{slides.json} format (containing only plain text), enabling LLM judges to perform direct pairwise comparisons between the full system and each baseline without formatting bias.

We employed \textbf{GPT-4o}\cite{openai2024gpt4o} as the primary judge to evaluate both \emph{Table Fidelity} (assessing the preservation of structured data, formulas, and precision) and \emph{Content Accuracy} (verifying factual and logical correctness), with \textbf{Claude-3.7-Sonnet}\cite{anthropic2024claude} serving as a replication model for fidelity tasks. To validate this LLM-as-judge protocol, two independent computer science Ph.D. students re-evaluated a random subset of papers; their judgments achieved a 90\% agreement rate with GPT-4o and high inter-rater reliability ($\kappa = 0.82$), confirming that the automated evaluations align strongly with domain-informed human consensus (full details in Appendix).

\textbf{Results.} 
Table~\ref{tab:results_summary} confirms that enhanced parsing yields substantial gains in multi-modal fidelity, while verification--adjustment provides supplementary robustness. Specifically, for \emph{Table Fidelity}, both judge models favored the Full system in 67.9\% of comparisons (vs.\ 17.9\% losses), underscoring the critical role of the Parser Agent in preserving complex structures. Conversely, \emph{Content Accuracy} was dominated by ties (85.7\%) with a marginal advantage for the Full system (7.1\% wins, 0\% losses), indicating that the base pipeline is inherently accurate and the Verification Agent acts primarily as a safeguard for borderline cases. The consistency across bidirectional comparisons and differing judge models reinforces the validity of these findings.

\begin{table}[t]
\centering
\caption{Results of the \textbf{LLM-as-judge} bidirectional ablation (Full system \emph{vs.}\ ablated variants). Win/tie rates are aggregated across forward and reverse orders. \emph{Agreement} denotes the consistency of judgments across positions.}
\begin{tabular}{llccc} 
\toprule
\textbf{Metric} & \textbf{Judge} & \textbf{Full Win} & \textbf{Tie} & \textbf{Agreement} \\
\midrule
\multirow{2}{*}{Table Fidelity} 
    & GPT-4o & 67.9\% (19) & 14.3\% (4) & 78.6\% \\
    & Claude & 67.9\% (19) & 14.3\% (4) & 50.0\% \\
\addlinespace 
Content Accuracy 
    & GPT-4o & \phantom{0}7.1\% (2)  & 85.7\% (24) & 71.4\% \\
\bottomrule
\end{tabular}
\label{tab:results_summary}
\vspace{-0.6cm}
\end{table}

\section{Conclusion}
We introduced \textsc{Auto-Slides}, an LLM-driven multi-agent framework that bridges scholarly writing and teaching materials by transforming papers into pedagogically optimized slide decks. By integrating high-fidelity multimodal parsing with narrative restructuring and verification loops, our system improves learner experience and strengthens structural support for understanding complex scientific content while preserving factual accuracy. This work advances the accessibility of academic research; detailed discussions on current limitations and future directions are provided in the Appendix.

\section*{Acknowledgement}\vspace{-0.6em}
This work was supported by the National Natural Science Foundation of China (No. 6250070674) and the Zhejiang Leading Innovative and Entrepreneur Team Introduction Program (2024R01007).
\vspace{-0.4em}

\bibliographystyle{IEEEbib}
\bibliography{ref}

\end{document}